\documentstyle[12pt,epsfig]{article}
\begin{document}
\begin{center}
{\bf
THE INFLUENCE OF STRONG INTERACTION ON THE PIONIUM WAVE FUNCTIONS AT
SMALL DISTANCES \\}
\vspace*{1cm} I. AMIRKHANOV, I. PUZYNIN, A. TARASOV, O.
VOSKRESENSKAYA \\and O. ZEINALOVA\\
{\it Joint Institute for Nuclear
Research, \\ 141980 Dubna, Moscow Region,\\ Russia}
\end{center}

\vspace*{.5cm}
\begin{abstract}
{\small The influence of strong $\pi^{+}\pi^{-}$-interaction of the
behaviour of pionium nS-state wave functions at small distance are
investigated both analytically (perturbatively) and so numerically. It is
shown that in the whole the accounting of strong interaction results in
multiplying of pure Coulomb pionium wave functions by some function
practically independent on value of principal quantum number n. Due to
this reason the n-independence of probability of $\pi^{+}\pi^{-}$-atom
production in nS-state remains the same as in case of pure Coulomb
$\pi^{+}\pi^{-}$-interaction.}
\end{abstract}

\vspace*{1cm}

The method of the measurement of pionium ($\pi^{+}\pi^{-}$-atom)
lifetime, proposed in the ref. \cite{C-1}, is essentially based on
the assumption that the n-dependence of probability $w_n$ of
creation of $\pi^{+}\pi^{-}$-atom in nS-state or is well known or, at
least, may be calculated with high degree of accuracy.

The first consideration of this problem have been done in the L. Nemenov's
paper \cite{C-2}, where the following relation
\begin{equation}
w_n\sim n^{-3}
\end{equation}
have been derived from more general result of author \cite{C-2}
\begin{equation}
w_n\sim\biggl\vert\int M(\vec r)\psi_n(\vec r)d^3r\biggl\vert^2\,,
\end{equation}
\begin{equation}
M(\vec r)=\frac{1}{(2\pi)^3}\int M(\vec p)e^{-i\vec p\vec r}d^3p\,,
\end{equation}
where $M(\vec p)$ being the amplitude of production of free
$\pi^{+}\pi^{-}$-pairs with relative momentum $\vec p$ in hadron --
hadron or hadron -- nucleus collisions and $\psi_n(r)$-being the
wave function of nS-state of pionium.

In his original derivation of (1) from (2) L. Nemenov besides accounting of
short range nature of amplitude M(r) also have made assumption that pure
Coulomb wave functions quite well describe distribution of pions in the
pionium not only at large distances, but at small ones also. However, as it
have been shown recently by E. Kuraev \cite{C-3}, this assumption is
unjustified due to noticeable influence of strong
$\pi^{+}\pi^{-}$-interaction on the behavior of pionium wave functions
in the nearest of origin. Due to this reason the more careful analysis of
this problem is needed.

Below we represent some preliminary results of such analysis, based on the
local potential model of strong $\pi^{+}\pi^{-}$-interaction.
In this model the "reduced" pionium wave functions
\begin{equation}
\Phi_n(r)=\sqrt {4\pi} r\psi_n(r), \quad \int\bigl\vert
\Phi_n(r)\bigl\vert^2dr=1\,,
\end{equation}
obey the Shr\"odinger equation
\begin{equation}
\Phi_n^{\prime\prime}(r)+m\bigl[U_c(r)+U_s(r)\bigl]\Phi_n(r)=
m\varepsilon_n\Phi_n(r)\,,
\end{equation}
where m being pion mass, $\varepsilon_n$ --- binding energy,
$U_c=\alpha/r$, $U_s$ --- Coulomb and strong potentials respectively.

First of all, let us apply the methods of perturbation theory to this
problem, treating the strong interaction potential $U_s$ as perturbation, in
order to obtain some qualitative estimations.
Putting
\begin{equation}
\Phi_n(r)\approx\Phi_n^{(0)}(r) + \Phi_n^{(1)}(r)\,,
\end{equation}
\begin{equation}
\varepsilon_n=\varepsilon_n^{(0)}+\varepsilon_n^{(1)}\,,
\end{equation}
where
\begin{equation}
\varepsilon_n^{(0)}=\frac{m\alpha^2}{4n^2}, \quad \varepsilon_n^{(1)}=
\int\limits_0^{\infty}U_s(r)\Phi_n^{(0)2}(r)dr\,;
\end{equation}
\begin{equation}
\Phi_n^{(0)\prime\prime}(r)+m\bigl[U_c(r)-\varepsilon_n^{(0)}\bigl]
\Phi_n^{(0)}(r)=0\,,
\end{equation}
\begin{equation}
\Phi_n^{(1)\prime\prime}(r)+m\bigl[U_c(r)\varepsilon_n^{(0)}\bigl]\Phi_n^
{(1)}(r)=
m\bigl[\varepsilon_n^{(1)}-U_s(r)\bigl]\Phi_n^{(0)}(r)\,;
\end{equation}
and applying the general methods of the solving of linear inhomogeneous
equations [4, 5] (see also \cite{C-6}) one can obtain
\begin{equation}
\Phi_n^{(1)}=\Phi_n^{(0)}\left[c_n-\chi_n(r)\right]\,,
\end{equation}
where
\begin{equation}
\chi_n(r)=\int\limits_{0}^{r}\frac{dr_1}
{\bigl\vert \Phi_n^{(0)}(r_1)\bigl\vert^2}
\int\limits_{0}^{r_1}\bigl\vert \Phi_n^{(0)}(r_1)\bigl\vert^2
\bigl[-\varepsilon_n^{(1)}+U_s(r_2)\bigl]dr_2
\end{equation}
and
\begin{equation}
c_n=\int\limits_{0}^{\infty}dr\bigl\vert \Phi_n^{(0)}(r_1)\bigl\vert^2
\chi_n(r)\,.
\end{equation}

If we define the ratio
\begin{equation}
R_n(r)=\frac{\Phi_n(r)}{\Phi_n^{(0)}(r)}\equiv
\frac{\psi_n(r)}{\psi_n^{(c)}(r)}
\end{equation}
as measure ot the influence of strong interactions on the values of the
pionium wave functions, then in the first order of perturbation theory
\begin{equation}
R_n(0)\approx 1+c_n\,.
\end{equation}
With explicit expressions for the pure Coulomb wave functions $\Phi_n^{(0)}$
we have proceeded in the calculation of $c_n$ with $n=1,\:2,\:3.$

The result looks like following:
\begin{equation}
c_n=\int\limits_{0}^{\infty}mU_s(r)rdr+\sum_{n=1}d_k^{(n)}\int
\limits_{0}^{\infty}mU_s(r)r\biggl(\frac{r}{r_B}\biggl)^k\ln
\biggl(\frac{r_k^{(n)}}{r}\biggl)dr\,,
\end{equation}
where $d_k^{(n)}\sim 1$, $r_k^{(n)}\sim r_B\approx 400\:fm$.

Taking into account the relation
\begin{equation}
\int\limits_{0}^{\infty}mU_s(r)r^2dr\approx a\approx 0.15\:fm~,
\end{equation}
where $a$ being $\pi^{+}\pi^{-}$ scattering length, it is easily to see that
n-dependent contributions to $c_n$ are numerically small (of order
$10^{-3}$) and may be neglected.

Putting $U_s=g/r\exp (-br)$ with values of parameters \cite{C-3}
$g\approx 3$,  $b=m_{\rho}\approx 3.8\:fm^{-1}$, that corresponds of applying
of the $\rho$-exchange model for describing of
strong $\pi^{+}\pi^{-}$-interaction, one can obtain for n-independent
part of $c_n$
\begin{equation}
\int\limits_{0}^{\infty}mU_s(r)rdr=\frac{gm}{m_{\rho}}\approx
0.55\,.
\end{equation}

These estimations show that the strong $\pi^{+}\pi^{-}$-interaction can
noticeably change the value of pionium wave functions at origin and this
effect can't be ignored at evaluating of probabilities $w_n$ (2). On the
other hand the large correction to the values at wave functions, obtained in
the first order of perturbation theory, means that the higher order
corrections are not small and must be taken into account. The calculation of
these ones is not simple problem. Due to this reason we have applied
numerical methods for accurate investigation of behaviour pionium wave
functions at small distances. In order to calculate the normalized pionium
wave functions we have used improved version of code \cite{C-7}, based on
applying of continuous analog of Newton method, developed in
the ref.\ \cite{C-8}

The input parameters of the code have been chosen in such a way, that
guaranteed accuracy of calculations was better than $10^{-4}$. To check this
accuracy we have compared numerical solution of Shr\"odinger equation
with pure Coulomb interaction with analytical one.

The results of numerical solution of equation (5) with Yukawa-type strong
potential, confirmed the main conclusions of perturbative consideration,
namely: the ratios
\begin{equation}
R_n(r)=\frac{\psi_n(r)}{\psi_n^{(c)}(r)}
\end{equation}
being numerically large (see Fig.\ 1) in the region $r\leq r_s\sim 1\:fm$,
essential for the problem under consideration (n-dependence of values
$w_n$), are practically n-independent (their n-independence is illustrated
by Fig.\ 2).

This means that with high degree of accuracy one can substitute
\begin{equation}
\psi_n(r)=R(r)\psi_n^{(c)}(r)
\end{equation}
in eq.\ (2) and, replacing $M(\vec r)\Rightarrow \widetilde M(\vec r)=M(\vec
r)R(r)$, obtain
\begin{equation}
w_n\sim\biggl\vert\int\widetilde M(\vec r)\psi_n^{(c)}dr\biggl\vert^2\sim
\biggl\vert \psi_n^{(c)}(0)\biggl\vert^2\biggl\vert\int\widetilde M(\vec
r)dr\biggl\vert^2\sim n^{-3}\,.
\end{equation}

\newlength{\pict}
\pict = 0.48\textwidth

\noindent
\parbox[t]{\pict}{
\mbox{\epsfig{file=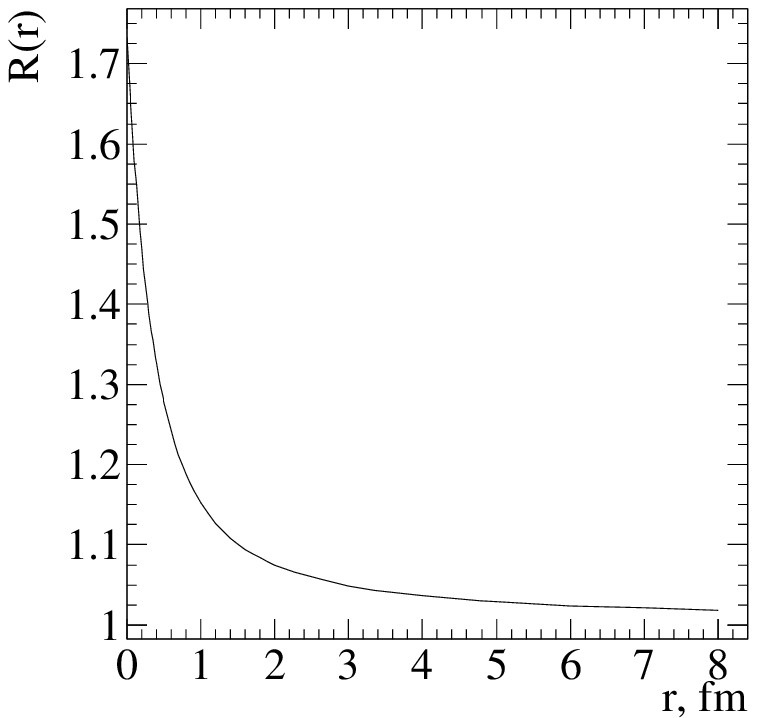,width=\pict}}

\refstepcounter{figure}
\label{f22r}
Figure\enskip\ref{f22r}: Influence of strong pion-pion interaction on
the pionium wave functions at small distances.}
\hfill
\parbox[t]{\pict}{
\mbox{\epsfig{file=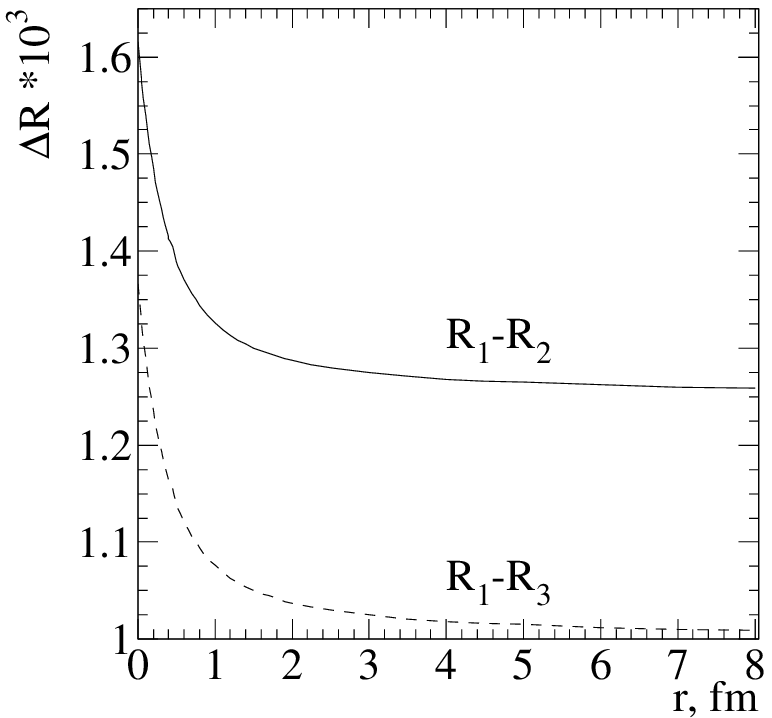,width=\pict}}
\refstepcounter{figure}
\label{fjj1}
Figure\enskip\ref{fjj1}: The accuracy of n-independence of strong
renormalization factors.}
\vspace*{0.5cm}

Thus we can conclude, that strong interaction corrections to pionium
wave functions at small distances, being sufficiently large, don't change
$n^{-3}$-law (1), primarily derived in the paper \cite{C-2} under assumption,
that pionium wave functions are pure Coulomb.

Authors are sincerely grateful to T. Pyzinina and E. Zemlaynaya for their
code.
Supported by the RFBR Grants No 97--01--01040, 97--02--17612.

\end{document}